\documentclass[twocolumn,tighten]{aastex62}

\usepackage{graphicx}
\usepackage{bm}
\usepackage{url}
\usepackage{threeparttable}
\usepackage{amsmath,amssymb}
\usepackage{xspace}
\usepackage{ulem}
\usepackage{subfigure}

\renewcommand\thefootnote{\roman{footnote}}

\newcommand{\nustar}{{\it NuSTAR}\xspace}
\newcommand{\xmm}{{\it XMM-Newton}\xspace}

\newcommand{\ergs}{erg~s$^{-1}$}
\newcommand{\gpa}{J0749+3337\xspace}
\newcommand{\gpb}{J0822+2241\xspace}

\newcommand{\modtext}[1]{\textbf{\textcolor{blue}{#1}}}

\shorttitle{A \nustar and \xmm Study of GPs} 
\shortauthors{Kawamuro et al.}

\begin{document}
\title{
  A \nustar and \xmm Study of the Two Most Actively Star-forming Green Pea Galaxies \\  
  (SDSS J0749+3337 and SDSS J0822+2241) 
} 

\correspondingauthor{Taiki Kawamuro}
\email{taiki.kawamuro@nao.ac.jp}

\author{Taiki Kawamuro}
\altaffiliation{JSPS fellow}
\affil{National Astronomical Observatory of Japan, Osawa, Mitaka, Tokyo 181-8588, Japan}

\author{Yoshihiro Ueda} 
\affil{Department of Astronomy, Kyoto University, Kitashirakawa-Oiwake-cho, Sakyo-ku, Kyoto 606-8502, Japan}

\author{Kohei Ichikawa}
\affil{Frontier Research Institute for Interdisciplinary Sciences, Tohoku University, Sendai 980-8578, Japan}

\author{Masatoshi Imanishi}
\affil{National Astronomical Observatory of Japan, Osawa, Mitaka, Tokyo 181-8588, Japan}

\author{Takuma Izumi}
\altaffiliation{NAOJ fellow}
\affil{National Astronomical Observatory of Japan, Osawa, Mitaka, Tokyo 181-8588, Japan}

\author{Atsushi Tanimoto}
\altaffiliation{JSPS fellow}
\affil{Department of Astronomy, Kyoto University, Kitashirakawa-Oiwake-cho, Sakyo-ku, Kyoto 606-8502, Japan}

\author{Kenta Matsuoka}
\altaffiliation{JSPS fellow}
\affil{Dipartimento di Fisica e Astronomia, Universit\'a degli Studi di Firenze, Via G. Sansone 1, I-50019 Sesto Fiorentino, Italy}
\affil{INAF – Osservatorio Astrofisico di Arcetri, Largo Enrico Fermi 5, I-50125 Firenze, Italy}

\begin{abstract}

  We explore X-ray evidence for the presence of active galactic nuclei (AGNs) 
  in the two most actively star-forming Green Pea galaxies (GPs), 
  SDSS~\gpa and SDSS~\gpb, which have star-formation rates (SFRs)
  of $123~M_\odot$~yr$^{-1}$ and $78~M_\odot$~yr$^{-1}$, respectively. The GPs have red 
  mid-infrared (MIR) spectral energy distributions and higher 22~$\mu$m luminosities
  than expected from a proxy of the SFR (H$\alpha$ luminosity), consistent with
  hosting AGNs with 2--10 keV luminosities of $\sim 10^{44}$ \ergs. We thus obtain 
  and analyze the first hard ($>$ 10 keV) X-ray data observed with \nustar and 
  archival \xmm data below 10 keV. From the \nustar $\approx$20~ksec data, 
  however, we find no significant hard X-ray emission. By contrast, soft X-ray emission
  with 0.5--8 keV luminosities of $\approx10^{42}$ erg~s$^{-1}$ is significantly
  detected in both targets, which can be explained only by star formation (SF). 
  A possible reason for the lack of clear evidence is that a putative AGN torus absorbs most of the 
  X-ray emission. Applying a smooth-density AGN torus model, we determine 
  minimum hydrogen column densities along the equatorial plane ($N_{\rm H}^{\rm eq}$)
  consistent with the non-detection.
  The results indicate $N_{\rm H}^{\rm eq}   
  \gtrsim 2\times10^{24}$~cm$^{-2}$ for SDSS~\gpa and $N_{\rm H}^{\rm eq} \gtrsim 5\times10^{24}$~cm$^{-2}$
  for SDSS~\gpb. Therefore, the GPs may host such heavily obscured AGNs. Otherwise, no AGN 
  exists and the MIR emission is ascribed to SF. Active SF in low-mass galaxies
  is indeed suggested to reproduce red MIR colors. This would imply that diagnostics
  based on MIR photometry data alone may misidentify such galaxies as AGNs.

  
\end{abstract}


\section{INTRODUCTION}\label{sec:int}

There now seems to be a general consensus that supermassive black holes (SMBHs)
with masses above a million solar masses \citep[$M_\odot$; ][]{Kor95,Kor13}
are ubiquitous in the center of massive galaxies. The growth history of SMBHs  
can be traced based on the luminosity functions of active galactic nuclei \citep[AGNs;
  e.g., ][]{Ued03,Ued14,Sha04,Has05}, and the results suggest that mass 
accretion is a dominant mechanism. This further infers the existence of massive black holes 
(mBHs) with masses in the range of $\sim 10^{3-6}~M_\odot$  \citep[][]{Mar04}. 
Given the correlation between the central SMBH mass and stellar properties of the
galaxy bulge \citep{Mag98,Geb00,Mar03,Gul09}, mBHs are predicted to reside in
low-mass galaxies and, indeed, have been found observationally \citep{Tho08,Bal15,Ngu17,Ngu18}. 

Some theoretical studies, however, have argued that not all low-mass galaxies host mBHs 
\citep{Vol08,Vol10} and that it depends on seed formation mechanisms such as a remnant 
of massive Population III stars \citep{Bro11}, the end-product of very massive 
stars formed through stellar mergers in dense star clusters \citep[e.g.,][]{Gur04},
and the result of the direct collapse of primordial dense gas \citep{Hae93,Beg06,Lod06}.
In other words, the mBH occupation fraction, as well as the mBH mass function in local 
low-mass galaxies, are expected to provide insights into how the seeds of SMBHs formed. 
Thus, mBH fractions have been enthusiastically measured under the assumption 
that the observed fractions of AGNs should be independent of the galaxy mass.
\citep{Gre12,Rei16c,Mez16,Mez18}. So far, fractions constrained 
using soft X-ray observations \cite[e.g., ][]{Gre12} support the view that direct collapse 
is a dominant process, where a lower occupation fraction is expected. Given that 
heavily obscured AGNs may be missed, however, a higher fraction is possible and may indeed 
favor the other scenarios. Thus, in order to draw a robust conclusion, it is necessary 
to construct as unbiased a sample as possible.

X-ray surveys are very important for sample construction \citep[e.g., ][]{Che17}.
As described above, the soft X-ray ($<$ 10 keV) band has often been used for such studies but
is easily biased against obscured systems. Moreover, given a theoretical prediction of 
increased soft X-ray luminosity in young and low-metallicity galaxies \citep[e.g., ][]{Lin10,Fra13}
and subsequent soft X-ray observations that have confirmed this \citep{Bas13,Bro16,Bro17}, 
it is possible to misidentify star-formation-induced soft X-ray emission as that
from an AGN. By contrast, the hard X-ray ($>$ 10 keV) band overcomes the above difficulties due to 
its high penetrating power and reduced contamination by stellar light. 
Mid-infrared (MIR) color-color selection is another option that is unbiased
against absorption; it relies on characteristic MIR colors produced by AGN-heated hot
dust \citep{Jar11,Ste12,Mat12} and has been examined for various samples \citep[e.g., 
][]{Gan15,Kaw16b,Ich17}. Some studies applied Wide-field Infrared Survey Explorer
(\textit{WISE}) AGN diagnostics to low-mass galaxies and created large AGN candidate
samples \citep{Sat14,Sar15,Sec15}. However, \cite{Hai16} demonstrated that star-forming
low-mass galaxies, particularly those with very young stellar populations and high
specific star-formation rates (sSFRs), could produce MIR colors similar to those of
\textit{WISE}-selected AGN. Thus, hard X-ray data are important and need to be investigated.

In this paper, we discuss the presence of AGNs in two low-mass galaxies ($M_\star \sim 10^{9}~M_\odot$) 
SDSS J074936.77+333716.3 and SDSS J082247.66+224144.0 (hereafter, \gpa and \gpb), i.e., the two highest
star-formation rate (SFR) Green Pea galaxies (GPs). As explained below, they 
  are optically classified as non-AGN hosts, but their MIR properties are consistent with having AGNs. 
Their basic properties can be found in Table~\ref{tab:inf_gps}. 

Through the Galaxy Zoo project \citep{Lin08}, GPs were first identified in the local Universe ($0.1 < z < 0.4$) 
by their green, unresolved (i.e., $\lesssim 1$ arcsec) compact morphology in Sloan Digital Sky Survey (SDSS) 
images \citep{Car09}. These features were interpreted as [O~III]$\lambda$5007 emission with high EWs
  ($\approx$700~\AA~on average) within $\approx$ 5 kpc. Figure~13 of \cite{Car09} demonstrated 
  that the EWs are generally higher than those observed in galaxies with similar redshifts and $g$-band magnitudes.
  Note that the EWs of \gpa and \gpb are $\approx$340~\AA~and $\approx$1040~\AA, respectively.
  \cite{Car09} reported that among 112 GPs with good quality optical spectra, nine GPs show broad Balmer lines, and
  thus were classified as AGN hosts. They applied the optical BPT diagram \citep{Kew01,Kau03} to the remaining 103
  sources, and the result was that 23 GPs are classified as AGNs while 80 GPs including our two targets as 
  star-forming galaxies.


The star-forming GPs have low stellar masses ($10^{8.5}$--$10^{10}$~$M_\odot$) and resemble 
high-redshift galaxies in terms of size, morphology, large emission lines, 
reddening, luminous UV emission (i.e., high SFRs), and low metallicity \citep{Car09,Izo11}. 
Thus, the GP sample is suggested to offer a valuable 
opportunity to investigate an early phase of galaxy growth in detail. Many interesting
results, such as star-forming conditions and the escape fraction of ionizing radiation,
have been reported, to date \citep[e.g., ][]{Cha12,Jas13,Hen15}. However, few studies have
mentioned the presence of AGNs by utilizing the MIR or soft X-ray observational data \citep[][]{Yan16,Svo18}.


With regard to our GPs, no AGN sign was found from optical spectra. 
  As suggested above, the GPs have inactive galaxy-like 
  [O~III]$\lambda$5007/H$\beta$ and [N~II]$\lambda$6583/H$\alpha$ flux
  ratios\footnote{We note that \gpb may have moderately high [O~III]$\lambda$5007/H$\beta$
    and [N~II]$\lambda$6583/H$\alpha$ flux ratios of 0.75 and -1.0 in logarithmic
    scale as calculated from the spectral line properties provided by the SDSS DR7 site
    of \texttt{http://skyserver.sdss.org/dr7/en/tools/search/radial.asp}.
    Thus, it may be classified as a AGN host, but in this paper we follow the results
    from a spectral analysis by \cite{Car09}. On the other hand, 
    the spectral lines of \gpa from the site are still consistent with an inactive galaxy}. 
  Also, their extinction-corrected [O~III] luminosity ($\approx 10^{43}$ erg s$^{-1}$) 
  to X-ray (2--10~keV) luminosity ratios, where the X-ray luminosities are estimated from
  the MIR emission (see Section~\ref{sec:wise_obs}), are slightly higher ($\approx$0.05) than
  the average of a nearby AGN sample of \cite{Ued15} ($\sim$0.03).
  The extinction correction is made by following \cite{Ued15}. 
  Thus, their [O~III] emission may be dominated by SF, consistent with the above. However, their MIR properties 
are consistent with those observed for AGN hosts (see Section~\ref{sec:wise_obs}). 
This apparent discrepancy could be explained if a mass accretion black hole is deeply 
  buried in the surrounding material and therefore the narrow line region remains absent \citep[e.g.,
  ][]{Ima01,Ima06,Ima08,Ima10,Ued07,Ich14}. To provide new insights into this discussion, we present the first hard 
X-ray data obtained with \nustar \citep{Har13}, currently the most sensitive hard X-ray (3--80 keV)
observatory. Additionally, soft X-ray properties are examined using \xmm \citep{Jan01} archival data. The
\nustar and \xmm observations were conducted in 2018 (PI: Kawamuro) and in 2013 (PI: Ehle),
respectively, and the log of these X-ray observations is given in Table~\ref{tab:xray_dat}.

\begin{deluxetable*}{ccccccccccccc}
\tabletypesize{\scriptsize}
\tablecaption{Information Relating to Two Green Pea Galaxies\label{tab:inf_gps}}
\tablewidth{0pt}
\tablehead{\vspace{-.1cm} \\ 
SDSS Name & R.A. (J2000) & Decl. (J2000) & $z$  & $D_{\rm L}$ & SFR &  $\log(M_\star/M_\odot)$ & $\log$(sSFR/Gyr$^{-1}$) & 12 + $\log $(O/H) & $L_{\rm H\alpha}$ \\ 
          & (degrees)    & (degrees)     &      & (Gpc) & ($M_\odot$ yr$^{-1}$) & & & & ($10^{42}$~\ergs) \\
      (1) &          (2) &          (3)  &  (4) & (5)   & (6)                   &  (7)   & (8) & (9) & (10) 
} 
\startdata
\gpa & 117.403215 & 33.621219 & 0.2733 & 1.40 & 123$\pm$51   & 9.49  & 1.11$\pm0.02$ & 8.29  & 7.4$\pm0.3$ \\
\gpb & 125.698590 & 22.695578 & 0.2163 & 1.07 & 78$\pm$34    & 8.43  & 1.98$\pm0.05$ & 8.08  & 4.7$\pm0.5$ 
\enddata 
\tablecomments{
  (1) SDSS source name. 
  (2) Right ascension. 
  (3) Declination. 
  (4) Redshift. 
  (5) Luminosity distance.
  (6) SFR that takes account of the dust-obscured and un-obscured SFRs (see Section~\ref{sec:dis2}
  for derivation). 
  (7) Stellar mass from \cite{Izo11}.
  (8) sSFR derived by following \cite{Hai16}, where SFRs were calculated with the H$\alpha$ luminosity as 
  $\log ({\rm SFR}/M_\odot~{\rm yr}^{-1}) = \log (L_{{\rm H}\alpha}/{\rm erg}~{\rm s}^{-1})
  - 41.27$.
  (9) Oxygen abundance ratio from \cite{Izo11}.  
  (10) Extinction-corrected H$\alpha$ luminosity taken from \cite{Car09}.  
 }
\end{deluxetable*}

The remainder of this paper is organized as follows. First, we briefly summarize the 
MIR properties of \gpa and \gpb based on the \textit{WISE} data in Section~\ref{sec:wise_obs}.
In Section~\ref{sec:nus_ana}, we present an analysis of the \nustar data, and report the
non-detection of the GPs in the hard X-ray band. In Section~\ref{sec:xmm_ana}, we
perform spectral analysis of the \xmm data. A discussion and summary are given in
Sections~\ref{sec:dis} and \ref{sec:sum}, respectively. Throughout this paper, we
assume a $\Lambda$CDM cosmology with $H_0=70$ km\,s$^{-1}$ Mpc$^{-1}$, $\Omega_{\rm m}=0.3$,
and $\Omega_{\Lambda}=0.7$. We utilize HEASoft version 6.22 for X-ray data reduction,
and XSPEC version 12.9.1p \citep{Arn96} for spectral analysis. The solar abundance 
table provided in \cite{Wil00}, where the oxygen abundance ratio of $12+ \log({\rm O}/{\rm H})$
is 8.69, is adopted. Errors attached to X-ray spectral parameters are given at 90\%
confidence limits for a single parameter of interest by following convention, while
others are quoted at the $1\sigma$ confidence level unless otherwise stated.

\section{\textit{WISE} MIR PROPERTIES}\label{sec:wise_obs}

\begin{figure}[!t] 
  \includegraphics[scale=0.48,angle=-90]{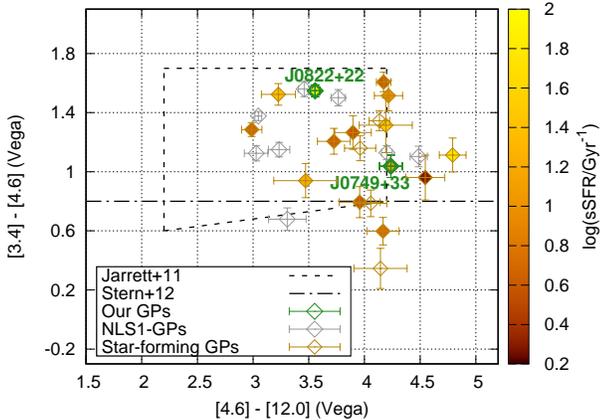}
  \caption{\small{
      MIR 4.6--12.0~$\mu$m versus 3.4--4.6~$\mu$m color-color plot. 
      Black dashed and dot-dashed lines represent the AGN selection criteria 
      proposed by \cite{Jar11} and \cite{Ste12}, respectively. 
      Only those whose stellar masses were constrained by \cite{Izo11} are color-coded 
      according to the right-hand color bar for the sSFR. 
      Four star-forming GPs are located outside the figure and are generally in the upper right direction. 
 }
 }\label{fig:mir_col} 
\end{figure}

  We present that in the MIR (3--22~$\mu$m) band the GPs have red colors, steep spectral
  indices, and luminous emission, consistent with the presence of an AGN.
Their MIR data were taken from the 
AllWISE data release, which combined the data taken from the four-band cryogenic phase \citep{Wri10} 
and the NEOWISE post-cryo phase \citep{Mai11}. The GPs were detected in all four bands
(W1: 3.4 $\mu$m, W2: 4.6 $\mu$m, W3: 12 $\mu$m, and W4: 22 $\mu$m) with S/N above 12
(i.e., \texttt{ph\_qual = A}) and little saturation (i.e., \texttt{w[1,2,3,4]sat}
$\approx$ 0). The photometry flag of \texttt{ccflag = 0} guaranteed that our sources
were unaffected by known artifacts (e.g., contamination and/or biased flux due to
proximity to an image artifact). The observed magnitudes were converted into flux densities 
by assuming a spectral index of $\alpha = 2$ in the form of $S_{\nu} \propto \nu^{-\alpha}$, 
close to those obtained by our spectral energy distribution (SED) fits (see below).

Figure~\ref{fig:mir_col} shows a \textit{WISE} color--color plot of the GPs together with two AGN
selection regions proposed by \cite{Ste12} and \cite{Jar11}: 
\begin{verbatim} 
 [3.4] - [4.6] >= 0.8 mag
\end{verbatim}
    and 
\begin{verbatim} 
 [4.6] - [12] > 2.2 mag 
 & [4.6] - [12] < 4.2 mag 
 & [3.4] - [4.6] > (0.1 x [4.6 - 12] + 0.38) mag 
 & [3.4] - [4.6] < 1.7 mag. 
\end{verbatim}
Here, we additionally add another 28 GPs that were detected in the four \textit{WISE} bands with
S/N $> 3$. The additional sample consists of 20 star-forming GPs and 8 AGN, or narrow-line Seyfert 
1 galaxies (NLS1), GPs, the details of which (i.e., R.A., and Dec.) are available in \cite{Car09}.
\gpa and \gpb satisfy both of the AGN criteria within uncertainty. All of the optically identified AGN
GPs can be classified as AGNs, and a large fraction of the star-forming GPs fall also within the criteria.
However, \cite{Hai16} suggested that the selections do not guarantee the presence of an AGN, 
particularly for low-stellar-mass, high SFR, or high sSFR ($\log({\rm sSFR}/{\rm Gyr}^{-1}) > 0.1$)
galaxies. Indeed, our GPs have high sSFRs (Figure~\ref{fig:mir_col})\footnote{By following \cite{Hai16},
  SFRs were calculated with H$\alpha$ luminosity as 
  $\log ({\rm SFR}/M_\odot~{\rm yr}^{-1}) = \log (L_{{\rm H}\alpha}/{\rm erg}~{\rm s}^{-1})
  - 41.27$.}.

MIR SEDs have often been used to identify AGNs by detecting a power law component originating 
in AGN-heated dust \citep[e.g., ][]{Pol07}. Spectral indices of luminous AGNs are typically 
$\alpha \gtrsim$ 0.5 in the form of $S_{\nu} \propto \nu^{-\alpha}$ \citep[][]{Alo06,Mul11}.
The spectral indices of \gpa and \gpb derived from single power law fits to W1, W2, and W4 band
photometry were $\alpha = 2.46\pm0.30$ and $2.02\pm0.21$, respectively. These are therefore
supportive of the presence of an AGN. The W3 band was excluded because various emission
(e.g., polycyclic aromatic hydrocarbon  (PAH) emission at the 7.7 $\mu$m, 8.6 $\mu$m, 11.3 $\mu$m,
and 12.7 $\mu$m bands) and absorption features (e.g., silicate absorption at 9.7 $\mu$m) contribute 
to emission. Note that we can obtain spectral indices consistent with those above 
even if we incorporate the W3 emission into the fits. 

We further investigate the origin of the MIR emission by focusing on the rest frame 22~$\mu$m 
  luminosities, which are $6.2\pm0.5\times10^{44}$ erg~s$^{-1}$ and $5.4\pm0.3\times10^{44}$ erg
  s$^{-1}$ for \gpa and \gpb, respectively. The luminosities were derived based on the W4 22-$\mu$m
band magnitudes of 6.8 and 6.4 in Vega for \gpa and \gpb and K-correction with $\alpha = 2$. 
  In discussing whether star formation (SF) is the 
only MIR source, we make a comparison with the expected SF luminosity. The W4 band is indeed
suited for this purpose because dust may cause emission at the shorter wavelengths (W1 and W2) to
become extinct, and also because 
a correlation between the W3 luminosity and SFR likely depends on the metallicity 
\citep[][]{Lee13}, making the comparison more complex. \cite{Lee13} derived a correlation
between the W4 band and H$\alpha$ luminosity, proxies of the SFR, as 
$\log (L_{{\rm H}\alpha}/{\rm erg~s^{-1}}) = 0.49 + 0.96 \log (L_{\rm W4}/{\rm erg~s^{-1}})$ 
with an intrinsic scatter of 0.21 dex. Their sample was composed of \textit{WISE} 
22~$\mu$m-detected star-forming galaxies at 0.01 $< z <$ 0.3 in an SDSS catalog. 
The H$\alpha$ luminosities, metallicities, and stellar masses of our GPs are within the
ranges considered in \cite{Lee13}. If all of the band W4 emission is ascribed to SF, 
  $\log (L_{{\rm H}\alpha}/{\rm erg~s^{-1}}) = 43.50\pm0.21$ and $43.43\pm0.21$ are predicted. 
These are significantly larger by $\approx$0.6--0.8~dex than those observed 
(42.87$\pm$0.02 of \gpa and 42.67$\pm$0.05 of \gpb). Thus, there may be an additional
contribution by AGNs. Conservative 22~$\mu$m AGN luminosities, from which the SF contribution 
expected by the H$\alpha$ emission is subtracted,
  are $4.8\pm0.9\times10^{44}$ \ergs and $4.5\pm0.6\times10^{44}$ \ergs for \gpa and \gpb, respectively. The AGN MIR luminosities 
can be converted into hard X-ray 14--150 keV luminosities of $\log (L_{\rm 14-150~keV}/{\rm erg}~{\rm s}^{-1})
= $ 44.70$\pm$0.48 and 44.67$\pm$0.47 with 1$\sigma$ scatters through the second equation in Table~3 of \cite{Ich17}
\footnote{Although the energy band represented in \cite{Ich17} is 14--195~keV, the correct one is
  14--150~keV.}. The correlation was derived based on \textit{Swift}/BAT hard X-ray-selected
nearby ($z < 0.3$) AGNs. Conventional 2--10 keV luminosities can be derived as 
  $\log (L_{\rm 2-10~keV}/{\rm erg}~{\rm s}^{-1}) = $ 44.29 and 44.26, respectively, by assuming
a cut-off power law with $\Gamma = 1.7$ and $E_{\rm cut} = 360$~keV.

\begin{figure}[!t] 
  \includegraphics[scale=0.7]{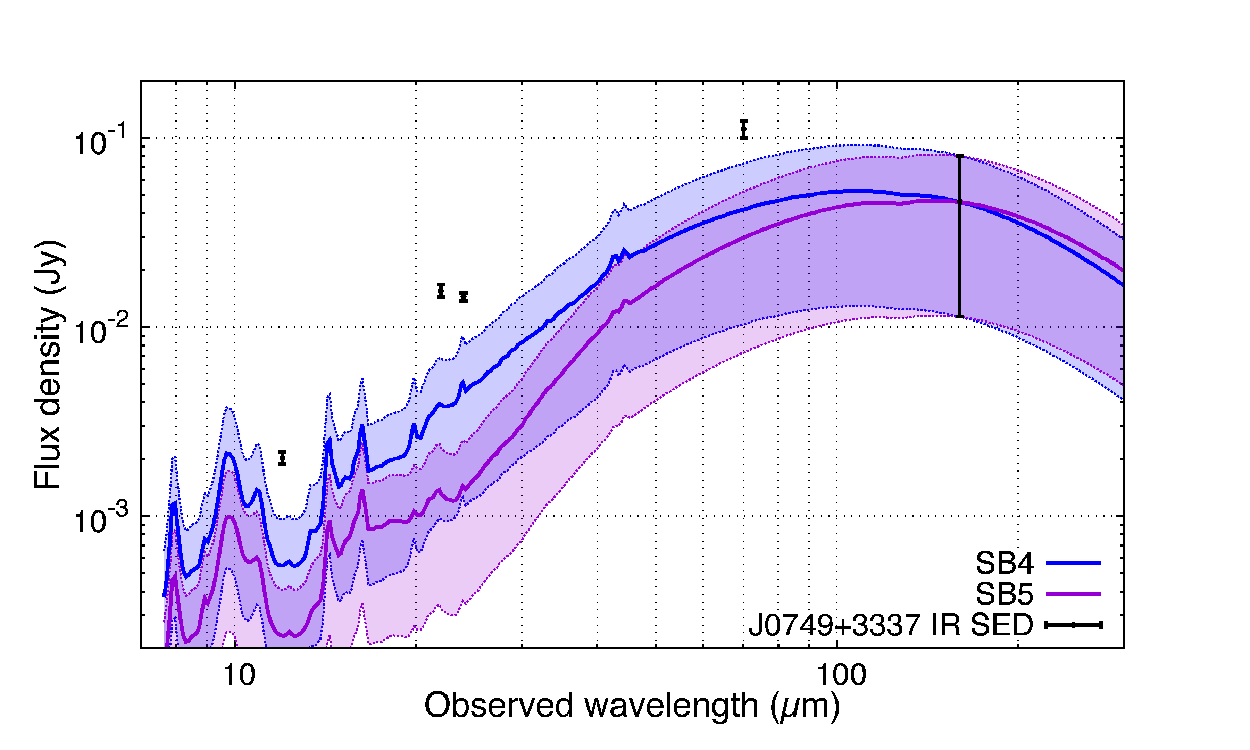}
  \caption{\small{
      IR SED of \gpa constructed from the \textit{WISE} and \textit{Spitzer}/MIPS
      data (black). Star-forming galaxy SEDs (SB4 and SB5 in \citealt{Mul11}),
      normalized at the 160~$\mu$m flux density, are represented by blue and magenta
      lines. The shades indicate regions enclosed by the model SEDs normalized
      at the 1$\sigma$ upper and lower 160~$\mu$m flux densities. 
 }
 }\label{fig:ir_sed} 
\end{figure}

For \gpa alone, we supplementarily examine the MIR luminosity expected from the far-infrared (FIR) luminosity
that traces the SFR. \gpb is not discussed here because no FIR data were available. We make a comparison 
between observed and model infrared (IR) SEDs (Figure~\ref{fig:ir_sed}). FIR (70~$\mu$m and 160~$\mu$m) 
and additional MIR 24~$\mu$m photometry data from the \textit{Spitzer}/MIPS is taken from \cite{Laa10}. 
The model SEDs are taken from \cite{Mul11}, who created five IR (6--1090~$\mu$m) templates by  
grouping 14 local ($< 80$ Mpc) star-forming galaxies in terms of their overall shape and the relative strength of 
their PAH features. Out of the five IR templates, we adopt two (SB4 and SB5 in \citealt{Mul11}) that considered  
galaxies with IR (8---1000~$\mu$m) luminosities of $\approx 10^{11.5}~L_\odot$, comparable to that of
\gpa \citep[$\sim 10^{11.7}~L_\odot$; ][]{Laa10}. Figure~\ref{fig:ir_sed} indicates MIR excess with respect 
to the models even in the most extreme case compatible with the observed 
160~$\mu$m flux densities. This is consistent with the above statement.

\section{\nustar Hard X-ray Data Analysis}\label{sec:nus_ana}

To obtain direct evidence for the presence of AGNs, we observed \gpa and
\gpb by \nustar, which carries two independent 
focal plane modules (FPMA and FPMB), with on-source exposures of $\approx$
19 ksec and $\approx$ 22 ksec, respectively. Following the ``\nustar Analysis
Quickstart Guide"\footnote{\texttt{http://www.srl.caltech.edu/NuSTAR\_Public/\\NuSTAROperationSite/SAA\_Filtering/SAA\_Filter.php}}, 
we used the standard \texttt{nupipeline} script for reprocessing. Our targets were 
very faint ($< 10^{-3}$ counts~s$^{-1}$), and periods of high background (such
as paths through or near the South Atlantic Anomaly (SAA)) must be excluded.
Typical background rates observed with \nustar are $\lesssim$ 1 count s$^{-1}$ 
integrated over the focal plane \citep{For14}. Times of high background can be
identified by simultaneously increased count rates in the detectors and shields that surround the focal planes. Using the
telemetry reports made by the \nustar team, we checked the total event rates during all orbital
passages of our observations. During the \gpa observation, the event rate slightly
increased around the standard SAA area ($\sim $ 2 counts s$^{-1}$). In addition,
high count rates occasionally occurred in the so-called tentacle region \citep{For14}
near the SAA. Thus, we ran \texttt{nupipeline} to reject times with high count rates
by setting options \texttt{saamode=optimized} and \texttt{tentacle=yes}. 
Background rates during the \gpb observation were stable and low, and thus we adopted 
\texttt{saamode=none} and \texttt{tentacle=no}. 

\begin{figure*}[!t] 
  \includegraphics[scale=0.6]{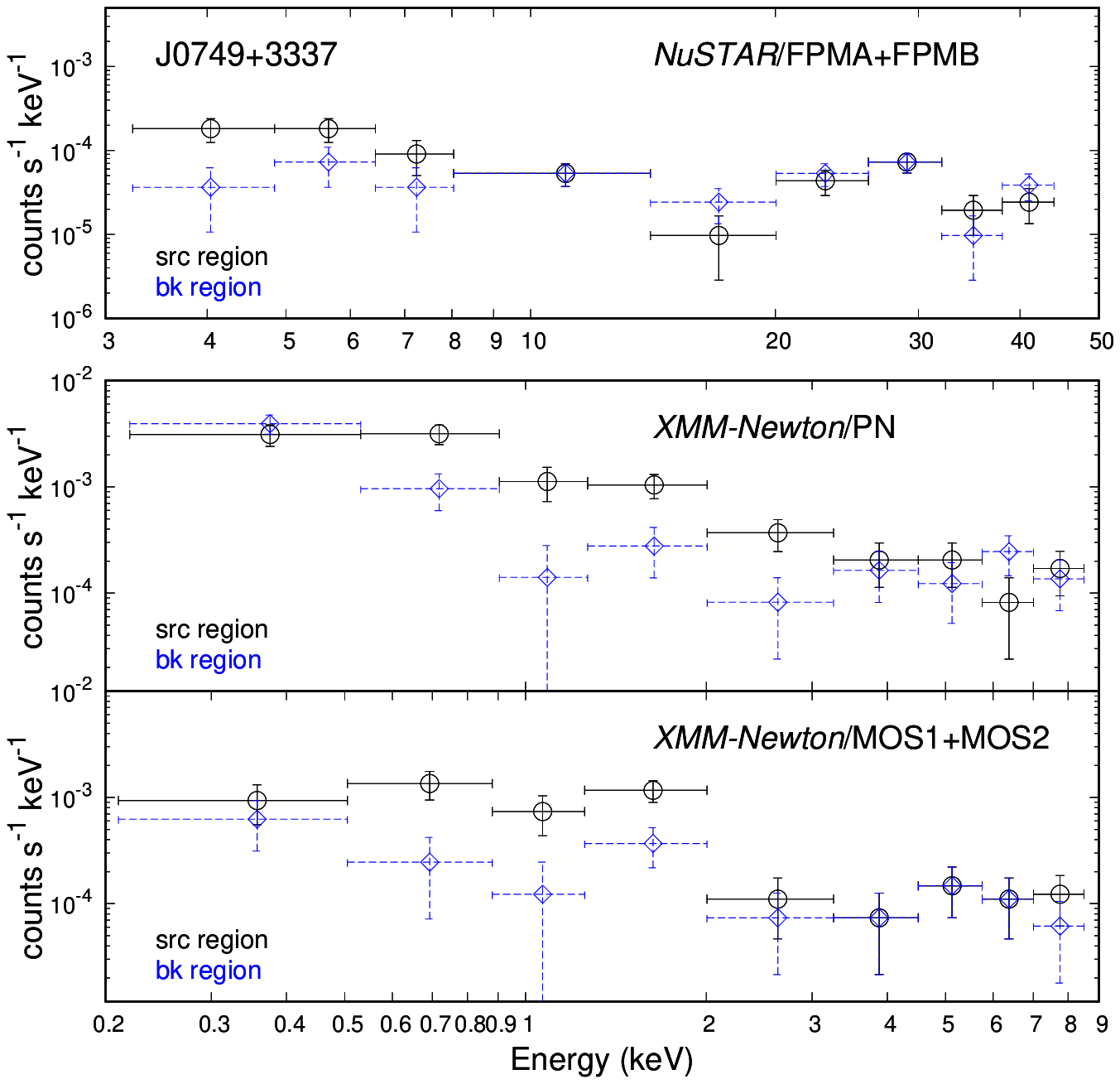}
  \includegraphics[scale=0.6]{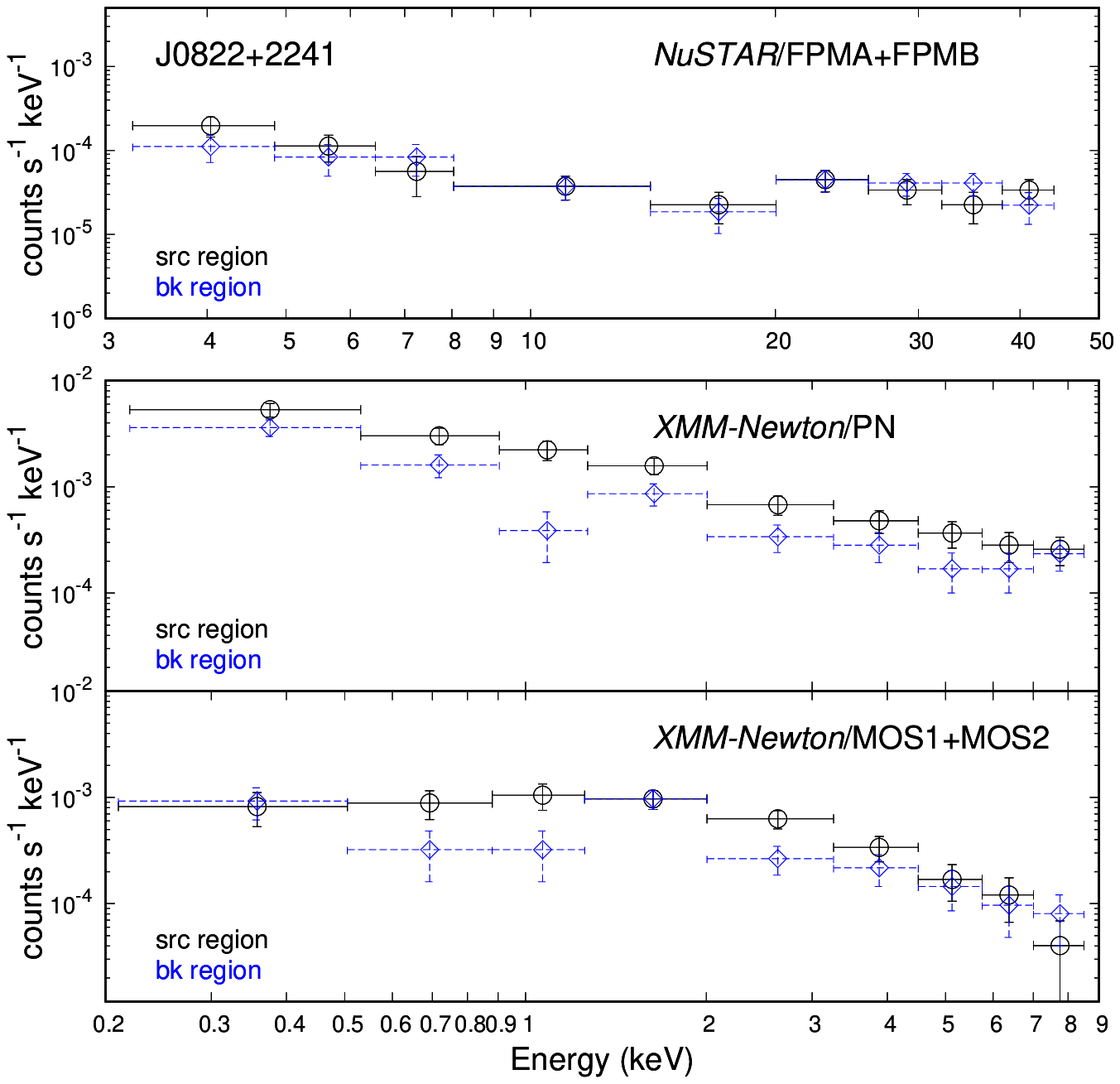}    \vspace{-1.3cm}
  \caption{\small{
 \nustar 3--50 keV and \xmm 0.2--9 keV spectra taken
      from the source (black crosses with circles) and background
      (blue dashed crosses with diamonds) regions. The figures
      suggest that the GPs are detected in the soft band, but
      not in the hard band. 
 }
 }\label{fig:nu_spe} 
\end{figure*}

We defined source regions as $30''$-radius circles centered at each optical position
by taking account of the full width at half maximum (FWHM) of the \nustar point spread
function (PSF; $\approx$18$^{\prime\prime}$). The size is much larger than the typical
size of GPs \citep[$< 1$ arcsec; ][]{Car09}. Background regions were off-source 
circular regions with a $30''$-radius on the same detector. Then, we produced source and 
background spectra, and response files using the \texttt{nuproducts} task. The 
products of FPMA and FPMB were combined to provide better statistics by using the 
\texttt{addascaspec} command. The systematic uncertainty between the two modules
is likely much smaller than the statistical uncertainty. Figure~\ref{fig:nu_spe} 
  shows the obtained spectra, as well as the background contribution. This clearly illustrates
  that we detected no significant emission from \gpa and \gpb even in the most sensitive
  8--24 keV band. Count rate upper limits at 3$\sigma$ in the energy range are $5.7\times10^{-4}$ 
  counts~s$^{-1}$ and $4.5\times10^{-4}$ counts~s$^{-1}$, respectively. 
  They are converted into 2--10 keV luminosities of $\approx 2\times10^{43}$ \ergs
  and $\approx 1\times10^{43}$ \ergs, respectively, by adopting a power law model with $\Gamma = 1.7$.
In Section~\ref{sec:dis1}, we further investigate how large absorbing column 
densities are needed to be consistent with the non-detection if there are 
  AGNs with $\log(L_{\rm 2-10~keV}/{\rm erg}~{\rm s}^{-1})\approx44.3$ (Section~\ref{sec:wise_obs}).

\begin{deluxetable}{cccccccccccccc} 
\tabletypesize{\footnotesize}
\tablecaption{X-ray Data List\label{tab:xray_dat}} 
\tablewidth{0pt}
\tablehead{
SDSS Name   & Observatory & ObsID & Obs. date (UT) & Exp.   \\
            &            &       &                & (ksec) \\ 
        (1) &        (2) &   (3) &            (4) & (5) 
}
\startdata
\gpa & \xmm    & 0690470101  & 2013 Mar. 25    & 20/22   \\
         & \nustar & 60301008002 & 2018 Mar. 23    & 34      \\ \hline 
\gpb & \xmm    & 0690470201  & 2013 Apr. 06    & 28/33   \\ 
         & \nustar & 60301009002 & 2018 Feb. 04 & 44   
\enddata
\tablecomments{
  Columns:
  (1) SDSS source name.
  (2) Observatory name. 
  (3) Observation ID. 
  (4) Observation start date. 
  (5) Exposure after data reduction. 
    For the \xmm observations, the PN and MOS 1 plus 2 exposure times are denoted separately, 
    while the FPMA and FPMB merged exposures are represented in the \nustar rows.   
}
\end{deluxetable}

\section{\xmm Soft X-ray Data Analysis}\label{sec:xmm_ana}

The \xmm data for \gpa and \gpb were obtained through the European Photon 
Imaging Camera/MOS (1 and 2) and PN detectors with duration time 
of $\approx$44 ks and $\approx$34 ks. All observations were performed by 
adopting the Prime Full Window mode and the thin filter. 

We reduced the data following the \xmm ABC guide\footnote{https://heasarc.gsfc.nasa.gov/docs/xmm/abc/}. 
The raw PN and MOS data were reprocessed using pipelines of 
\texttt{epchain} and \texttt{emchain}, respectively. To filter periods
with a high background, we created PN background light curves in the 10--12 keV band 
with \texttt{PATTERN = 0} (single events), and those of the MOS in energies above 10 keV with the same pattern selection. 
Regarding the \gpb data, we adopted background count
rate thresholds of 0.35 s$^{-1}$ and 0.40 counts s$^{-1}$ for the MOS and PN 
cameras, respectively. These are the recommended values in the guide. In contrast, 
we found two high-background flares during the \gpa observation, and could not clearly
remove the tails of the flares with the recommended thresholds. Thus, we excluded the 
first 10 ks and the last 11 ks to obtain clean data. The PN data were
further limited to those with \texttt{PATTERN $\leq$ 4} (single and double events) and
\texttt{FLAG = 0}, corresponding to the most conservative screening criteria. For the
MOS data selection, \texttt{PATTERN $\leq$ 12} (single, double, triple, and quadruple
events) was adopted. Central circular regions with $20''$ and $25''$ radii, larger 
than the FWHM of the \xmm PSF ($\approx 6''$), were set to extract \gpa and \gpb source events, 
respectively. The larger region was adopted for \gpb because its X-ray image seemed to be 
slightly extended, although this was likely due to low photon statistic fluctuation. Each
background spectrum was extracted from an off-source circular region with the same
radius as that used for the source events. The spectra from the MOS1 and
MOS2 detectors were combined into one. We analyzed the spectra in the 0.4--7.0 keV band, 
where \gpa and \gpb were significantly detected with S/N = (5.5, 4.1) and (6.6, 3.9) 
for (PN, MOS), respectively. The response files were 
generated in a standard manner for a point source. 

\subsection{\xmm Spectral Analyses}\label{sec:xmm_spe}

\begin{figure*}[!t]
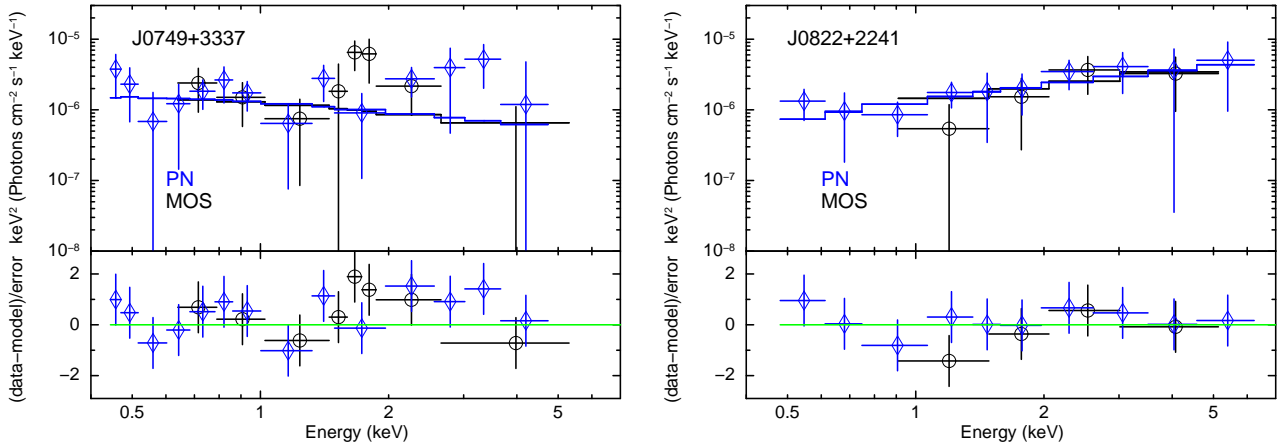
 
  \includegraphics[scale=0.33,angle=-90]{gp1.eps}
  \includegraphics[scale=0.33,angle=-90]{gp2.eps}
  \caption{\small{
      \xmm 0.4--7.0 keV spectra, corrected for the response functions. 
      Black crosses with circles and blue crosses with diamonds
      represent the MOS and PN spectra, respectively. 
      For clarity, the spectra are re-binned with larger bin sizes. 
      Each solid line is the best-fit un-absorbed power law model 
      (Model1 in Table~\ref{tab:xmm_fit}). Residuals are plotted
      in the lower panels.       
 }
 }\label{fig:xmm_spe} 
\end{figure*}

We simultaneously fit the PN and MOS spectra (Figure~\ref{fig:xmm_spe}) to increase the 
S/N. The spectra are binned so that each energy bin had at least one count. We thus
determine best-fit models based on the $C$-statistic \citep{Cas79}, appropriate for low
photon counts. Goodness of fit is examined by following the procedure given in \cite{Kaa17}, where 
the expected $C$-statistic value ($C_{\rm exp}$) and variance ($\sigma_{C_{\rm exp}}$)
from a model is compared with the observed
value ($C_{\rm obs}$). Note that the best-fit models found in the following sections are consistent
with the non-detection by \nustar. 

Essentially, we determine the best-fit models using the following model,
\begin{verbatim} 
 constant*TBabs*zTBabs*zpowerlw (Model1), 
\end{verbatim}
expressed in XSPEC terminology. The main component is the single absorbed power law
(\texttt{zTBabs*zpowerlw}), and is adopted for a consistent comparison with the work
by \cite{Bro16} (Section~\ref{sec:dis2}). They parameterized the X-ray emission of low-mass
galaxies by single power law fits and discussed its association with SF. The
power law component may be ascribed to emission from high-mass X-ray binaries (HMXBs)
and an AGN, if present. We also include the Galactic absorption, whose hydrogen column
density is estimated from the \texttt{nh} command in HEASoft \citep{Kal05}, 
with \texttt{TBabs}. We fix $N^{\rm Gal}_{\rm H} = 4.65\times10^{20}$ cm$^{-2}$
and $N^{\rm Gal}_{\rm H} = 4.13\times10^{20}$ cm$^{-2}$ for \gpa and \gpb,
respectively. To absorb systematic uncertainty in the 
normalization between the PN and MOS spectra, we apply the \texttt{constant}
model, whose value is represented by $C_{\rm MOS/PN}$. To avoid
implausible values, we allow it to vary only within 10\%, a canonical range
\citep[e.g., see Figures~6 and 7 of ][]{Mad17}.
We finally obtain four free parameters: the photon index ($\Gamma$) and normalization of \texttt{zpowerlw},
the absorbing column density ($N_{\rm H}$), and the cross-normalization between the spectra ($C_{\rm MOS/PN}$).
Errors in intrinsic luminosity from the power law are constrained by replacing \texttt{zpowerlw} with
\texttt{pegpwrlw}, which explicitly provides flux (or luminosity) errors in a given energy range.

We also apply another model that additionally takes into account emission 
from the hot interstellar medium (ISM) and young stellar objects (YSOs)
according to the \cite{Min12b}: 
\begin{verbatim} 
 constant*TBabs*zTBabs(zpowerlw+apec+zbremss) 
                                     (Model2). 
\end{verbatim}
The ISM emission is modeled by optically thin thermal emission (\texttt{apec}) with a temperature of 0.24
keV, the average value of those measured in nearby galaxies. The metal abundance is set to 0.40 for \gpa
and 0.25 for \gpb according to each oxygen abundance ratio. The YSO emission is modeled by bremsstrahlung 
(\texttt{zbremss}) with a canonical temperature of 3 keV \citep{Win07,Min12b}.
Luminosities from the ISM and YSOs are expected to increase with SFR as
$L_{\rm 0.5-2~keV, ISM}/{\rm SFR} = 5.2\times10^{38} $ (erg s$^{-1}$/$M_\odot$  
yr$^{-1}$) and $L_{\rm 2-10~keV, YSO}/{\rm SFR} = 1.7\times10^{38} $ (erg s$^{-1}$/$M_\odot$ yr$^{-1}$).
Accordingly, we fix the normalizations of the two thermal emission at those corresponding to the expected
luminosities. Note that we do not consider X-ray emission from low-mass X-ray binaries, cataclysmic
variables, or active 
binaries. Given correlations of their luminosities and stellar mass \citep{Gil04,Bog11}, at most 
$L_{\rm X} \sim 10^{38}$ erg s$^{-1}$ is expected from those populations. 
This is much smaller than observed luminosities ($\sim 10^{41}$ erg s$^{-1}$). Finally, we obtain the same four free 
parameters as in Model1. We stress that because the two models provide similar values (Table~\ref{tab:xmm_fit}), 
our discussion does not depend on the adopted models, as detailed below. 

\subsubsection{Soft X-ray Band \gpa Spectra}

Fitting Model1, we obtain an un-absorbed ($N_{\rm H} < 4.5\times10^{21}$ cm$^{-2}$), soft 
($\Gamma = 2.6^{+1.0}_{-0.8}$) power law model in $C_{\rm obs}$/$C_{\rm exp}$ = 92/87$\pm10$. 
The rest frame 0.5--8 keV intrinsic luminosity ($L_{\rm 0.5-8~keV}$) is measured 
to be 1.2$^{+0.9}_{-0.5}\times10^{42}$ erg s$^{-1}$. Model2 also provides a similar result, 
where $N_{\rm H} < 4.7\times10^{21}$ cm$^{-2}$, $\Gamma = 2.5^{+1.1}_{-0.}$, and
$L_{\rm 0.5-8~keV} =$ 1.2$^{+0.6}_{-0.5}\times10^{42}$ erg s$^{-1}$.

\subsubsection{Soft X-ray Band \gpb Spectra} 

Similarly to the \gpa case, Model1 can reproduce the \gpb spectra well with insignificant
absorption ($N_{\rm H} < $ 3.5$\times10^{21}$cm$^{-2}$) and a harder photon index 
($\Gamma$ = $1.3^{+0.8}_{-0.4}$) in $C_{\rm obs}$/$C_{\rm exp}$ = 200/210$\pm$15. The luminosity is 
$L_{\rm 0.5-8~keV} = 1.4\pm0.7\times10^{42}$ erg s$^{-1}$. A similar result can be obtained 
by fitting Model2 (Table~\ref{tab:xmm_fit}).

\begin{deluxetable*}{ccccccccccccccccc} 
\tabletypesize{\small}
\tablecaption{\xmm Spectral Analysis Results \label{tab:xmm_fit}} 
\tablewidth{0pt}
\tablehead{
  (1)  &  SDSS Name            &  & \multicolumn{2}{c}{\gpa} & \multicolumn{2}{c}{\gpb} \\
  (2) & Model & & Model1 & Model2 & Model1 & Model2 
}
\startdata
(3)  & $N^{\rm Gal}_{\rm H}$ & (10$^{20}$ cm$^{-2}$) & 
\multicolumn{2}{c}{4.65} & \multicolumn{2}{c}{4.13} \\ 
(4)  & $C_{\rm MOS/PN}$  &
& 0.97$^{+0.13}_{-0.07}$ & 0.96$^{+0.14}_{-0.06}$
& 0.98$^{+0.12}_{-0.08}$            & 0.98$^{+0.12}_{-0.08}$ \\
(5)  & $N_{\rm H}$           & (10$^{22}$ cm$^{-2}$)
& 0.00$(< 0.45)$ & 0.00$(< 0.47)$ 
& 0.00$(< 0.35)$ & 0.00$(< 0.38)$ \\
(6)  & $\Gamma$              &
& 2.6$^{+1.0}_{-0.8}$ & 2.5$^{+1.1}_{-0.9}$ 
& 1.3$^{+0.8}_{-0.4}$ & 1.3$^{+0.8}_{-0.5}$ \\ 
(7)  & Norm                  & (10$^{-6}$ photon cm$^{-2}$ s$^{-1}$ keV$^{-1}$)
& 2.6$^{+1.0}_{-0.9}$ & 2.3$^{+1.0}_{-0.9}$ 
& 1.9$^{+1.8}_{-0.6}$ & 1.8$^{+1.8}_{-1.0}$ \\ 
(8)  & $F_{\rm 0.5-2~keV}$   & (10$^{-15}$ erg cm$^{-2}$ s$^{-1}$) 
& 2.8 & 2.8 
& 3.1 & 3.1 \\ 
(9)  & $F_{\rm 0.5-8~keV}$   & (10$^{-15}$ erg cm$^{-2}$ s$^{-1}$) 
& 4.2 & 4.4 
& 11  & 11 \\ 
(10)  & $F_{\rm 2-10~keV}$    & (10$^{-15}$ erg cm$^{-2}$ s$^{-1}$) 
& 1.6 & 1.8 
& 10  & 10 \\ 
(11) & $L_{\rm 0.5-2~keV}$   & (10$^{41}$ erg s$^{-1}$) 
& 8.5$^{+5.5}_{-3.8}$ & 7.6$^{+3.7}_{-3.0}$ 
& 4.0$^{+5.5}_{-1.7}$ & 3.9$^{+3.5}_{-1.6}$ \\ 
(12) & $L_{\rm 0.5-8~keV}$   & (10$^{41}$ erg s$^{-1}$)
& 12$^{+9}_{-5}$ & 12$^{+6}_{-5}$ 
& 14$\pm7$       & 14$\pm5$ \\  
(13) & $L_{\rm 2-10~keV}$    & (10$^{41}$ erg s$^{-1}$)
& 4.2$^{+10.7}_{-3.5}$ & 4.5$^{+8.1}_{-3.4}$ 
& 13$\pm9$             & 13$\pm7$ \\ 
(14) & $C_{\rm obs}$/$C_{\rm exp}$/d.o.f &  
& 92/87$\pm$10/109  & 92/88$\pm$11/109 
& 200/210$\pm$15/241 & 200/210$\pm$15/241 \\ 
(15) & S/N (PN, MOS)         &
& \multicolumn{2}{c}{5.5, 4.1} & \multicolumn{2}{c}{6.6, 3.9} 
\enddata
\tablecomments{
  Columns:
  (1) SDSS source name.
  (2) Model1 is the absorbed power law, and Model2 consists of the absorbed 
  power law, the optically thin thermal emission, and bremsstrahlung. 
  The latter two thermal emission are fixed. 
  (3) Hydrogen column density of the Galactic absorption. 
  (4) Ratio between the MOS and PN spectral models. 
  (5) Hydrogen column density of extragalactic absorption. 
  (6) Power law photon index of the power law component.
  (7) Power law normalization at 1 keV. 
  (8)--(10) Observed fluxes in the 0.5--2, 0.5--8, and 2--10 keV bands. 
  (11)--(13) Absorption-corrected intrinsic luminosities in the 0.5--2, 0.5--8, and 2--10 keV bands. 
  (14) Observed $C$-statistic value, and expected $C$-statistic value with its 1$\sigma$ uncertainty,
  and degrees of freedom. 
  (15) S/N in the 0.4--7.0 keV band. 
  (The fluxes and luminosities are estimated from the PN spectra.)   
}
\end{deluxetable*}

\section{DISCUSSION}\label{sec:dis} 

\subsection{Origin of Soft X-ray Emission}\label{sec:dis2} 

We investigate the origin of the soft X-ray emission reported in Section~\ref{sec:xmm_spe} 
in terms of the luminosity. To discuss whether or not SF can reproduce $\approx$10$^{42}$
\ergs, we refer to \cite{Bro16}. They derived a correlation between the X-ray (0.5--8 keV)
luminosity, SFR, and oxygen abundance ratio (12 + $\log $(O/H)) for Lyman break analogues:
supercompact, UV-luminous galaxies at $z<$ 0.3, regarded as nearby analogues of
more distant Lyman break galaxies such as GPs. The SFR used in the correlation is defined as
the sum of dust-obscured and un-obscured SFRs \citep{Bro17} \cite[see also ][]{Hir03}.
The dust un-obscured SFRs of 59$\pm3$~$M_\odot$ yr$^{-1}$ for \gpa and 37$\pm4$~$M_\odot$
yr$^{-1}$ for \gpa derived by \cite{Car09} may therefore be underestimated. Thus, by
dividing the un-obscured H$\alpha$-based SFRs with an absorbed Lyman continuum fraction
of $0.48\pm0.20$ \citep[][]{Hir03}, we estimate the total SFRs to be 123$\pm$51~$M_\odot$
yr$^{-1}$ and 78$\pm34~M_\odot$ yr$^{-1}$ for \gpa and \gpb, respectively. 
Regarding the stellar mass and oxygen abundance ratio, we adopt those from \cite{Izo11}. 
The two values were derived, respectively, based on SED fits and the so-called direct method,
which uses the electron temperature within the [O~III] zone from the
[O~III]$\lambda 4363$/($\lambda 4959$+$\lambda 5007$) line ratio. Then, predicted 0.5--8~keV
luminosities from the correlation are $\log (L^{\rm SF}_{\rm 0.5-8~keV}/{\rm erg}~{\rm s}^{-1})
= 41.81\pm0.38$ (1$\sigma$) for \gpa and $\log (L^{\rm SF}_{\rm 0.5-8~keV}/{\rm erg}~{\rm s}^{-1})
= 41.74\pm0.39$ (1$\sigma$) for \gpb. These are consistent with the observed 0.5--8 keV
luminosities of $\log (L_{\rm 0.5-8~keV}/{\rm erg}~{\rm s}^{-1}) = 42.08^{+0.33}_{-0.18}$ 
for \gpa and $42.15\pm0.22$ for \gpb. Thus, the soft X-ray emission can be ascribed solely 
to SF, or likely HMXBs. Therefore, we cannot detect any evidence for soft X-ray emission from
AGNs.

\subsection{\nustar Non-detection due to Heavy Obscuration?}\label{sec:dis1}

The 22 $\mu$m luminosities of our sample would seem to originate from AGN
emission, and their expected luminosities would be $\log(L_{\rm X}/{\rm erg~s^{-1}})
\sim 44$ (see Section~\ref{sec:wise_obs}). However, \nustar does not show any 
significant detection from those sources. This seemingly contradictory result
could be naturally described if the central engine is heavily obscured. Mainly
utilizing the \nustar 8--24 keV data, we compute how large absorbing hydrogen
column densities are required to be consistent with the non-detection.

We construct our model by adopting a Monte-Carlo-based numerical AGN torus model 
\footnote{The model is available from
  \\ \texttt{https://heasarc.gsfc.nasa.gov/xanadu/xspec/models/etorus.html}}\texttt{e-torus}. 
The original version was created by \cite{Ike09} and has often been used to study 
AGN tori \citep[e.g., ][]{Taz13,Kaw13,Kaw16b,Ric13,Ric14,Gua16,Tan16,Tan18,Oda18,Yam18}.
The \texttt{e-torus} model calculates reflected spectra from constant-density cold matter that has two 
cone-shaped holes along the polar axis \citep[see Figure~2 of][]{Ike09}. The ratio of 
the inner and outer radii is fixed at 0.01. The solar metal abundance is adopted. The torus
property is determined by the hydrogen column density in the equatorial plane ($N_{\rm H}^{\rm eq}$),
and the half opening ($\theta_{\rm op}$) and inclination ($\theta_{\rm inc}$) angles. 
These angles are defined so that 0$^\circ$ and 90$^\circ$ correspond to the pole direction and 
the equatorial direction, respectively. The GPs are optically non-active galaxies, implying that
a torus with a small opening angle prevents formation of the narrow line region \citep[i.e., ][]{Ued07}.
Thus, within the acceptable range of $\theta_{\rm op} = 10^\circ$--$70^\circ$,
  we adopt $\theta_{\rm op}$ = 10$^\circ$, corresponding to a covering factor of 98\%, 
while $\theta_{\rm inc}$ is set to 60$^\circ$ as a representative value to ensure an 
obscured AGN. The primary X-ray emission is modeled by a cut-off power law
spectrum with a high-energy cut-off of 360 keV and $\Gamma = 1.7$ \citep[e.g., ][]{Kaw16a}.
  Adopting alternative values (i.e., $\Gamma = 1.9$, $\theta_{\rm inc} = 89^\circ$, and
$\theta_{\rm op} = 37^\circ$) instead of the default values, we can confirm that our 
conclusion is little affected. Specifically, our model is represented as 
\begin{verbatim} 
 torusabs*zpowerlw*zhighect 
  +zpowerlw*zhighect
   *mtable{e-torus_20161121_2500M.fits} 
  +atable{refl_fe_torus.fits} 
  +[Best-fit Model of the XMM-Newton Spectra], 
\end{verbatim}
almost the same as those used in past studies \citep[e.g., ][]{Tan18,Oda18}. From the
first to the third terms, we take account of the absorbed cut-off power law component from 
an AGN, the reflected emission, and the accompanying 6.4 keV iron-K$\alpha$ line. The fourth
term means that we include emission expected from each best-fit model (Model1) determined in 
the \xmm spectra (Section~\ref{sec:xmm_spe}). 

\begin{figure*}[!t] 
  \includegraphics[scale=0.71]{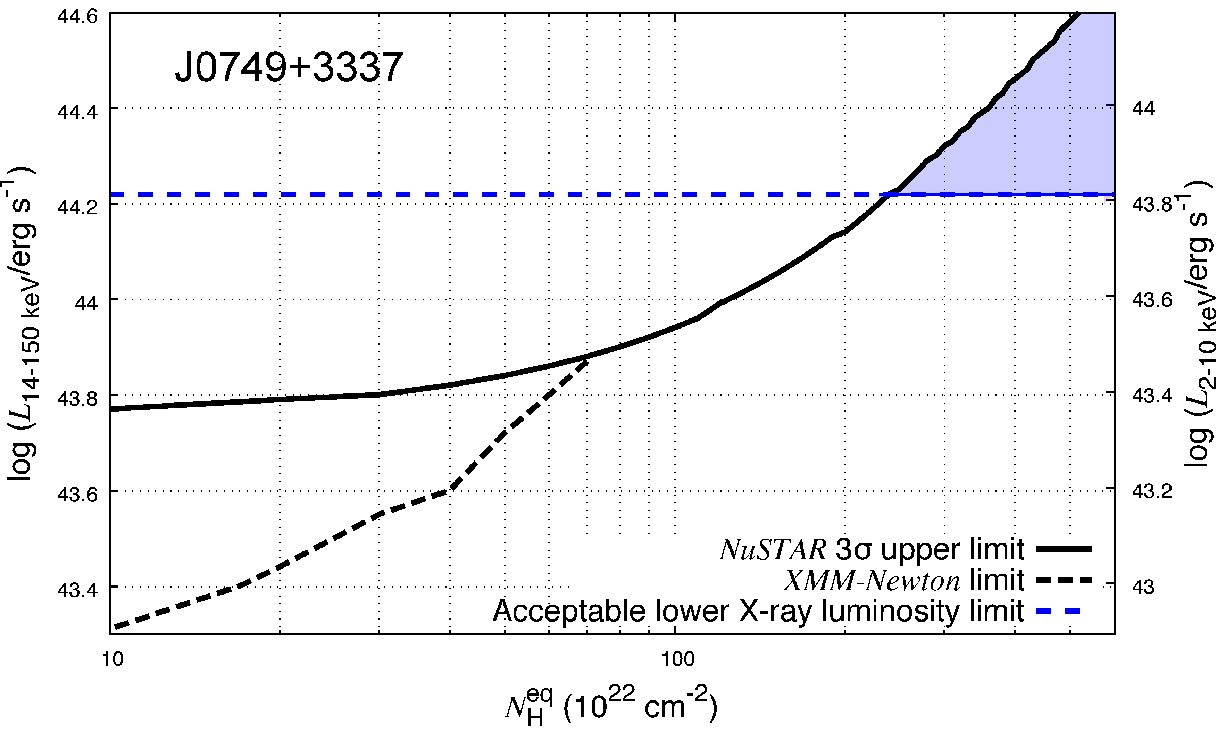}
  \includegraphics[scale=0.71]{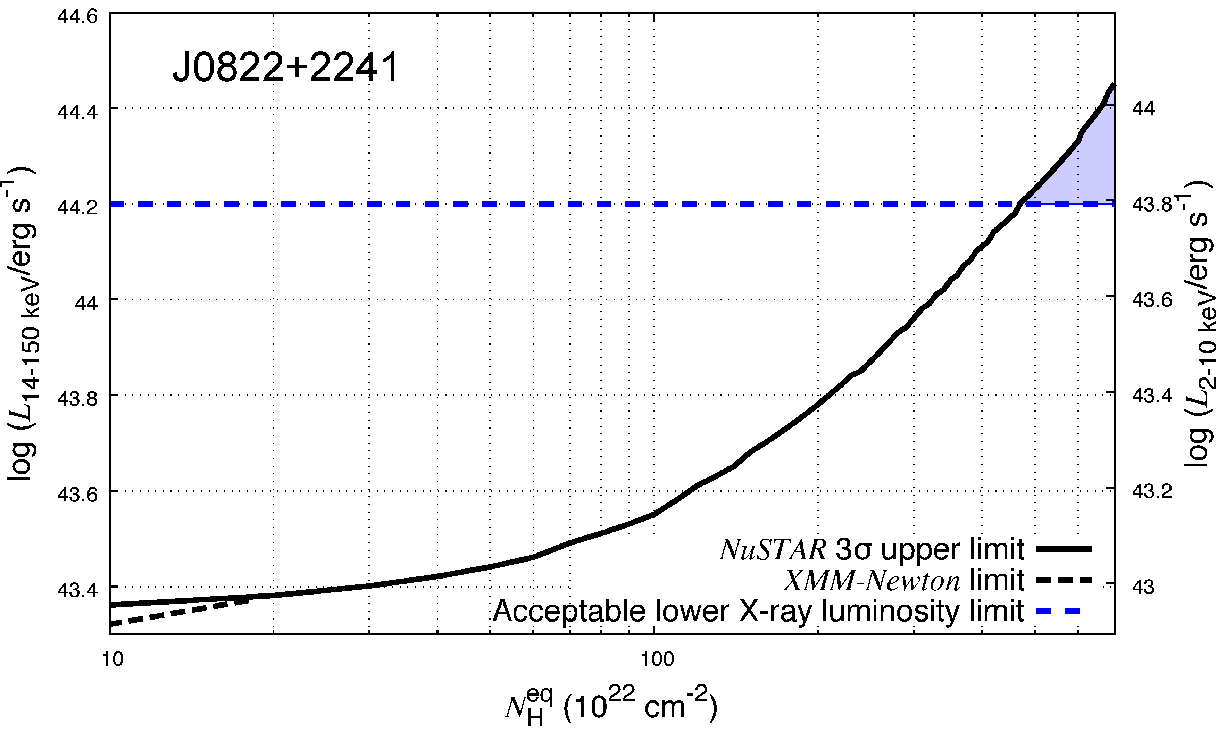}  
  \caption{\small{
      Expected X-ray luminosity versus hydrogen column density in the equatorial plane. 
      The right $L_{\rm 2-10~keV}$-axis is scaled with the left $L_{\rm 14-150~keV}$-axis
      under a cut-off power law model with $\Gamma = 1.7$ and $E_{\rm cut}$ = 360 keV.
      The blue dashed line represents the lower X-ray luminosity 
      limit expected from the observed 22~$\mu$m luminosity for each source. The maximum 
      luminosities accepted by the \xmm spectra for a given column density are denoted
      by black dashed lines. In other words, luminosities above the limits erroneously
      exceed the \xmm spectra. Blue shades correspond to acceptable areas and indicate 
      lower limits of $N^{\rm eq}_{\rm H} \gtrsim 2\times10^{24}$ cm$^{-2}$ and 
      $N^{\rm eq}_{\rm H} \gtrsim 5\times10^{24}$ cm$^{-2}$ for \gpa and \gpb.
 }
 }\label{fig:nh_upp} 
\end{figure*}

We estimate power law normalizations that reproduce the 3$\sigma$ source count rates for
various column densities in the equatorial plane ($N_{\rm H}^{\rm eq}$), and compute
corresponding intrinsic X-ray luminosities. The result is plotted in Figure~\ref{fig:nh_upp}
and is compared with the 14--150~keV (plus 2--10~keV) luminosities expected from the 22 $\mu$m
ones to constrain acceptable ranges of $N_{\rm H}^{\rm eq}$. A point of concern is that
when the column density is lower than a certain value, the torus models tend to exceed the
\xmm spectra while being consistent with the \nustar observations. Therein, we take account
of the maximum luminosity accepted by the \xmm spectra and also plot the results. Specifically,
we fit a suite of torus models having a given column density with various normalizations
together with an absorbed power law model to the \xmm spectra. Then, we compute $C_{\rm obs}$,
$C_{\rm exp}$, and $\sigma_{C_{\rm exp}}$ following \cite{Kaa17}. Finally, we search for a
maximum normalization where $C_{\rm obs} = C_{\rm exp} + 2.78\times \sigma_{C_{\rm exp}}$, 
equivalent to 3$\sigma$, and plot the corresponding luminosity. Eventually,
  Figure~\ref{fig:nh_upp}
indicates that the column density ($N_{\rm H}^{\rm eq}$) must be larger than $2\times10^{24}$
cm$^{-2}$ for \gpa and $5\times10^{24}$ cm$^{-2}$ for \gpb.
Thus, if present, their AGNs should be heavily obscured.

The low metal abundances of the GPs make photoelectric absorption more ineffective compared
with solar absorption. In this case, higher column densities are needed. 
As a simple estimate, by considering that the column density is inversely proportional to the metal
abundance for a given level of absorption,
  $N_{\rm H}^{\rm eq}\gtrsim 5\times10^{24}$ cm$^{-2}$ for \gpa and
  $N_{\rm H}^{\rm eq}\gtrsim 2\times10^{25}$ cm$^{-2}$ for \gpb are expected.
The estimate may be reasonable but un-rigorous for various reasons; for example, the reflection
component seen around 30~keV and the Compton scattering are not taken into 
consideration in this discussion.

\subsection{MIR Emission due to SF?}\label{sec:dis3} 

We also discuss another possibility for a non-AGN case for our GPs. 
In this case, the red MIR colors and steep spectral slopes may be ascribed to YSOs. 
Note that the asymptotic giant branch (AGB) star is another stellar MIR 
emitter, but is not likely to be the main source given that the W3-W4 colors 
of our GPs (3.6 and 2.9 for \gpa and \gpb) are redder than expected from
usual AGBs \citep[$\lesssim$2; ][]{Koe14,Lia14}. It has been suggested
that very young YSOs in particular, with an age $\lesssim$ a few Myr
show NIR and MIR emission from optically thick disks
\citep[e.g., ][]{Lad87,Str89,Hai01,Dun14}. Motivated by this fact, 
some studies proposed selection and classification criteria for YSOs
that use the WISE data \citep[e.g.,][]{Koe14,Kan17}. 
The results indicate that Class I and II YSOs have MIR colors 
similar to those observed in luminous AGNs; that is, in our GPs as well.
Moreover, a MIR index defined as $d \log(\lambda S_\lambda)/d\log \lambda$ 
has often been used for the classification of conventional YSO classes.
\citep{Gre96,Mar13,Maj13,Kan17}. Characterizing the MIR slopes of our GPs 
based on \cite{Mar13}, we find that they show 
$d \log(\lambda S_\lambda)/d\log \lambda \approx 1$, consistent with 
those of Class I YSOs. Thus, if SF is the main source that powers the 
MIR emission, the Class I YSO is a plausible type of star that mainly
contributes to it. This YSO interpretation is consistent with the idea
that GPs correspond to an early phase of galaxy formation.

\section{Summary} \label{sec:sum}

To discuss whether or not the two GPs (\gpa and \gpb) host AGNs, suggested 
from the MIR \textit{WISE} observations, we obtained the initial hard X-ray
($>$ 10 keV) data using \nustar. Then, including the \xmm data ($<$ 10 keV), 
we explored X-ray evidence for the presence of an AGN. Our results are
summarized as follows. 

\begin{itemize}
\item Both GPs were detected in all \textit{WISE} bands, and have red colors,
  steep spectral indices of $\alpha \sim 2$, and higher MIR luminosities 
  ($\approx 5$--$6\times10^{44}$ erg~s$^{-1}$) than expected 
  from the H$\alpha$ emission. These data are consistent with the presence of an AGN. 
\item We detected no significant hard X-ray (8--24 keV) emission from the GPs.
\item Soft (0.4--7 keV) X-ray emission was significantly ($> 3\sigma$) detected.
  The 0.5--8 keV luminosities reach $\approx10^{42}$ erg~s$^{-1}$, and can be
  explained by SF only. 
\item Considering the AGN X-ray luminosities expected from the MIR data, we estimated the 
  minimum column densities required to be consistent with the non-detection
  by \nustar.
  The result indicated that if present, the AGNs in \gpa and \gpb 
  were obscured with column densities 
  $N^{\rm eq}_{\rm H}\gtrsim 2\times10^{24}$ cm$^{-2}$ and 
  $N^{\rm eq}_{\rm H}\gtrsim 5\times10^{24}$ cm$^{-2}$, respectively.
  If smaller abundances were assumed, larger column densities of
  $N^{\rm eq}_{\rm H}\gtrsim 5\times10^{24}$ cm$^{-2}$ for 
  \gpa and $N^{\rm eq}_{\rm H}\gtrsim 2\times10^{25}$ cm$^{-2}$ for \gpb were expected.
\item Finally, the possibility remains that no AGN exists and is not the main source that powers the
  MIR emission. In this case, young Class I YSOs would be plausible main contributors in the MIR band 
  (Section~\ref{sec:dis3}). This is consistent with a previous study  (Section~\ref{sec:int}) where
  low-mass galaxies with active SF, such as GPs, probably in an early phase of galaxy growth, were 
  suggested to reproduce photometric MIR properties similar to those of AGN hosts. If true,
  this implies that diagnostics that use MIR photometry data alone may misidentify such
  galaxies as AGNs. 
\end{itemize}

To finally distinguish between the two possibilities, MIR spectroscopy may be an 
option. Detection of a deep silicate absorption feature at 9.7 $\mu$m may favor 
the presence of an AGN deeply embedded in the dust \citep[e.g., ][]{Dud97,Eva03}, 
because it is difficult to produce if sources are largely distributed, as in SF. 
Also, equivalent widths of PAH emission may be examined given that a lower (higher) 
value is expected if an AGN (SB) dominates the MIR emission \citep[e.g., ][]{Ima07,Vei09}.
  In the future, the \textit{JWST}/MIRI spectroscopy observation with 
  high sensitivity will be a promising way to draw a strong conclusion.
  
\acknowledgments

This work was financially supported by the Grant-in-Aid for
Japan Society for the Promotion of Science (JSPS) Fellows
(T.K., K.M., and A.T.), and for Scientific Research (KAKENHI)
17K05384 (Y.U.), 18K13584 (K.I.), 15K05030 (M.I.), and 17K14247 (T.I.).
K.I. was further supported by Program for Establishing a Consortium
for the Development of Human Resources in Science and Technology,
Japan Science and Technology Agency (JST). Also, this study 
was based on data from the NuSTAR mission, a project led by the
California Institute of Technology, managed by the Jet Propulsion
Laboratory and funded by NASA, and also observations obtained
with XMM-Newton, an ESA science mission with instruments and
contributions directly funded by ESA Member States and NASA.
This research made use of the NuSTAR Data Analysis Software
(NuSTARDAS) jointly developed by the ASI Science Data Center
and the California Institute of Technology.

\end{document}